\documentclass[12pt,a4paper,final]{iopart}

\usepackage{graphicx}
\usepackage{dcolumn}
\usepackage{bm}
\usepackage{hyperref,color}

\newcommand{\ad}{\mathrm{(ad)}}
\newcommand{\nad}{\mathrm{(nad)}}
\newcommand{\eq}{\mathrm{(eq)}}

\begin{document}

\title{Linear stochastic thermodynamics for periodically driven systems}
 
\author{Karel Proesmans}
 \ead{Karel.Proesmans@uhasselt.be}
 \address{Hasselt University, B-3590 Diepenbeek, Belgium.}
\author{Bart Cleuren}
\address{Hasselt University, B-3590 Diepenbeek, Belgium.}%
\author{Christian Van den Broeck}
\address{Hasselt University, B-3590 Diepenbeek, Belgium.}%

\date{\today}

\begin{abstract}
The theory of linear stochastic thermodynamics is developed for periodically driven systems in contact with a single reservoir. Appropriate thermodynamic forces and fluxes are identified, starting from the entropy production for a Markov process. Onsager coefficients are evaluated, the Onsager-Casimir relations are verified, and explicit expressions are given for an expansion in terms of Fourier components. The results are illustrated on a periodically modulated two level system including the optimization of the power output.
\end{abstract}


\section{Introduction}
Over the past decade, thermodynamic properties of small systems have been studied extensively, culminating in the development of stochastic thermodynamics \cite{harris_fluctuation_2007,sekimoto2010stochastic,seifert_stochastic_2012,SpinneyFord,van_den_broeck_ensemble_2014,tome2015stochastic}. 
This theory allows to perform a detailed and in many respects novel thermodynamic analysis of nonequilibrium steady states, complementing the long list of results obtained for such systems 
in the context of nonequilibrium statistical mechanics, see for example \cite{mclennan1959statistical,oono1998steady,hatano2001steady}. 
However, thermodynamic processes often operate via cycles or time-periodic processes, which function as engines of transformation between different flows of energy. The Carnot cycle is the prime example. The stochastic thermodynamics of such periodically driven systems has also been investigated, both theoretically \cite{jung1993periodically,PhysRevB.83.153417,esposito_finite-time_2010,lahiri2012fluctuation,chirla2014finite,rana2014single,PhysRevE.92.042108,proesmans2015efficiency,holubec2015asymptotics,dare2015time} and experimentally \cite{blickle2006thermodynamics,martinez_brownian_2014,jun2014high,koski2014experimental,martinez2015adiabatic,rossnagel2015single}. Following some revealing case-studies \cite{schmiedl2008efficiency,izumida2009onsager,esposito2010quantum,izumida2010onsager,izumida2012efficiency,izumida2015linear}, important steps towards a general formulation, akin to the classical formulation of linear irreversible thermodynamics \cite{callen2006thermodynamics,kondepudi2014modern}, have been made recently \cite{brandner2015thermodynamics,proesmans2015onsager}.
The purpose of the present paper is to further develop this theory by presenting the linear stochastic thermodynamic theory for a Markovian system under
 time-periodic variation of unspecified control parameters. A key step is the identification of proper thermodynamic fluxes and forces, obtained by averaging over one period, leading to the standard bi-linear form of the entropy production and to the Onsager-Casimir relation for the corresponding Onsager coefficients. An expansion into Fourier components reveals the underlying symmetry properties of the Onsager matrix.
 
 The outline of the paper is as follows.
 To set the scene, we develop in section \ref{Gce} the linear stochastic thermodynamics for a periodically driven system in contact with a heat and particle reservoir with modulated temperature and chemical potential. These results inspire the formulation, in section \ref{Gt}, of the general theory for a system in contact with a single reservoir, but with arbitrary periodic modulation of an otherwise unspecified set of control parameters. In section \ref{fouran}, we show how the thermodynamic forces and fluxes are to be expanded in terms of the Fourier components, revealing the bare underlying symmetry structure of the Onsager matrix. In section \ref{exa}, the results are applied to a single level quantum dot in contact with an electron reservoir. Work can be delivered by appropriate variation of the temperature of the reservoir and of the energy of the quantum dot. Finally, we close with a brief discussion and outlook in section \ref{disc}.

\section{Grand canonical ensemble \label{Gce}}
Before developing the general theory, it is revealing to consider a periodically driven system in contact with a single reservoir characterised by a time-dependent temperature $T(t)$ and chemical potential $\mu(t)$. This will also serve to introduce the basic set-up and notation. For simplicity we consider a system described by a set of discrete states $m$, with corresponding energy $\epsilon_m(t)$ and number of particles $n_m$. The energy of any particular state can change in time due to the interaction with a work source, and transitions between states can take place by exchanging heat and particles with the reservoir. The state of the system is described by the probability distribution $\boldsymbol{p}(t)=\{p_m(t)\}$, with $p_m(t)$ the probability to be in state $m$ at time $t$. The dynamics is assumed to be Markovian, i.e., the time evolution is described by a master equation:
\begin{equation}
\dot{\boldsymbol{p}}(t)=\boldsymbol{W}(t)\boldsymbol{p}(t).\label{MEq}
\end{equation}
$\boldsymbol{W}(t)$ is the transition matrix. 
In absence of any modulation, the transition matrix reduces to a time-independent expression, denoted by $\boldsymbol{W}^{\eq}$. The corresponding steady state $\boldsymbol{p}^{\eq}$, $\boldsymbol{W}^{\eq}\boldsymbol{p}^{\eq}=\boldsymbol{0}$, reproduces the equilibrium grand canonical distribution:
\begin{equation}
p^{\eq}_m=\frac{1}{\mathcal{Z}^{eq}}{e^{-\frac{\epsilon_{0,m}-\mu_0n_m}{T_0}}};\;\;\; \mathcal{Z}^{eq}=\sum_m e^{-\frac{\epsilon_{0,m}-\mu_0n_m}{T_0}},
\end{equation}
where ${\epsilon}_{0,m}$, $T_0$ and $\mu_0$ are the reference values for the energy level $m$, temperature and chemical potential. For the sake of notational simplicity, we have set the Boltzmann constant $k_B=1$ throughout the text.

\subsection{Thermodynamic forces}
The energies of the system states are modulated by an external source as follows:
\begin{equation}\label{ForceDef3}
\frac{\epsilon_m(t)}{T_0}=\frac{\epsilon_{0,m}}{T_0}+F_{\epsilon}{\gamma_{\epsilon,m}}g_{\epsilon}(t),
\end{equation}
where $F_{\epsilon}$ is a measure for the strength of the driving. The factor $\gamma_{\epsilon,m}$ accounts for the relative amplitude at which the different levels $m$ are modulated. $g_{\epsilon}(t)$ describes the corresponding time dependence, which we take here to be uniform over all energy levels.
In the same way, the modulation of temperature and chemical potential can be represented as:
\begin{eqnarray}
\frac{1}{T(t)}&=&\frac{1}{T_0}+F_Tg_T(t),\label{ForceDef1}\\
\frac{\mu(t)}{T_0}&=&\frac{\mu_0}{T_0}+F_\mu g_\mu(t).\label{ForceDef2}
\end{eqnarray}
In the following, we consider time-periodic driving with a single common period $\mathcal{T}$, hence all time-dependent driving functions $g_{\alpha}(t)$, $\alpha \in \{\mu,T,\epsilon\}$ satisfy:
\begin{equation}
g_{\alpha}(t+\mathcal{T})=g_{\alpha}(t).
\end{equation}
These modulations bring the system out of equilibrium and induce heat, chemical and mechanical work fluxes. Our main purpose is to identify the corresponding entropy production, in the long time limit and averaged over one period, as a bi-linear function of the appropriate thermodynamic forces and fluxes, in full analogy with standard linear irreversible thermodynamics.

To obtain a correct expression for the entropy production, it is crucial to incorporate the following "self-consistency conditions" on the time-dependent transition matrix. Consider the "instantaneous" steady state distribution $\boldsymbol{p}^{\ad}(t)$, satisfying:
\begin{equation}
\boldsymbol{W}(t)\boldsymbol{p}^{\ad}(t)=0,\label{dbdef}
\end{equation}
at every instance of time $t$. We refer to this distribution as the 'adiabatic' distribution, as it is the solution of the master equation for adiabatically slow driving. Since the system is assumed to be in contact with a single reservoir, this distribution at time $t$ has to be the equilibrium distribution with the driving parameters fixed at their instantaneous values:
\begin{equation}p^{\ad}_m(t)=\frac{1}{\mathcal{Z}}e^{-\frac{\epsilon_m(t)-\mu(t)n_m}{T(t)}};\;\;\;
\mathcal{Z}=\sum_me^{-\frac{\epsilon_m(t)-\mu(t)n_m}{T(t)}}.\end{equation}
In addition, the adiabatic distribution has to satisfy the following detailed balance condition:
\begin{equation}
W_{nm}(t)p^{\ad}_{m}(t)=W_{mn}(t)p^{\ad}_{n}(t).
\end{equation}
In analogy with standard irreversible thermodynamics, these conditions could be referred to as "local equilibrium".

It is clear that the parameters $F_\alpha$, describing the amplitudes of the driving, are good candidates for the thermodynamic forces.
As the focus here is on linear thermodynamics, i.e.~small thermodynamic forces, we can approximate the adiabatic distribution by its linear expansion:
\begin{equation}
p^{\ad}_m(t)\approx p^{\eq}_m-F_{\mu}\delta\gamma_{\mu,m}g_\mu(t)-F_{T}\delta\gamma_{T,m}g_T(t)-F_{\epsilon}\delta\gamma_{\epsilon,m}g_{\epsilon}(t),\label{pdbgce}
\end{equation}
for all $m$. Here, we introduced a uniform notation:
\begin{eqnarray}\gamma_{\mu,m}&=&-n_m,\;\;\;\gamma_{T,m}=\epsilon_{0,m}-\mu_0n_m,\label{gamdef}
\end{eqnarray}
and
\begin{equation}\label{deltagamma}
\delta\gamma_{\alpha,m}=p^{\eq}_m\bigg(\gamma_{\alpha,m}-\sum_k p^{\eq}_k{\gamma}_{\alpha,k}\bigg).
\end{equation}
For further reference, we notice that $\sum_m\delta \gamma_{\alpha,m}=0$, due to the normalisation of both $\boldsymbol{p}^{\eq}$ and $\boldsymbol{p}^{\ad}(t)$.

\subsection{Thermodynamic fluxes}
To identify the thermodynamic fluxes, we now turn to the expressions for work and heat. Even though the system is periodically modulated in time, we can identify a steady operation as follows. In the long time limit, the system returns after each period to the same statistical state. By performing averages over one cycle, we effectively operate in a "steady state" regime. The corresponding averaged quantities will be denoted by an overbar. We first turn to the mechanical work. According to the theory of stochastic thermodynamics \cite{seifert_stochastic_2012,van_den_broeck_ensemble_2014}, work corresponds to the change in energy of an occupied state of the system. The rate of work (or power), averaged over one period, can thus be written as follows:
\begin{eqnarray}\dot{\overline{W}}&=&\frac{1}{\mathcal{T}}\int^{\mathcal{T}}_0dt\, \dot{W}(t)\nonumber\\&=&\frac{1}{\mathcal{T}}\int^{\mathcal{T}}_0dt\,\sum_m\dot{\epsilon}_m(t)p_m(t)\nonumber\\&=&\frac{F_{\epsilon}T_0}{\mathcal{T}}\int^{\mathcal{T}}_0dt\sum_{m}{\gamma}_{\epsilon,m}\dot{g}_{\epsilon}(t){p_m}(t)\nonumber\\&=&\frac{F_{\epsilon}T_0}{\mathcal{T}}\int^{\mathcal{T}}_0dt\sum_{m}\delta{\gamma}_{\epsilon,m}\dot{g}_{\epsilon}(t)\frac{{p_m}(t)}{p^{\eq}_m},\end{eqnarray}
or more compactly:
\begin{eqnarray}\label{mechwork}
\dot{\overline{W}}&=&T_0F_\epsilon J_\epsilon,\\
J_\epsilon&=&\frac{1}{\mathcal{T}}\int^{\mathcal{T}}_0dt\,\langle \delta{\boldsymbol{\gamma}}_{\epsilon}\vert \boldsymbol{p}(t) \rangle \dot{g}_{\epsilon}(t).
\end{eqnarray}
Here $J_\epsilon$ is identified as the work flux, and $\langle.|.\rangle$ is the following inner product:
\begin{equation}\left\langle \boldsymbol{k}|\boldsymbol{l}\right\rangle=\sum_m \frac{k_m l_m}{p^{\eq}_m}.\label{inpdef}\end{equation}
As a result of the detailed balance condition, this inner product has a symmetry with respect to the transition matrix at zero thermodynamic forces $\boldsymbol{W}^{\eq}$:
\begin{equation}\left\langle \boldsymbol{k}|\boldsymbol{W}^{\eq}\boldsymbol{l}\right\rangle=
\left\langle \boldsymbol{W}^{\eq}\boldsymbol{k}|\boldsymbol{l}\right\rangle.
\end{equation}

The chemical work can be defined in an analogous way:
\begin{eqnarray}\dot{\overline{W}}_{chem}&=&\frac{1}{\mathcal{T}}\int^{\mathcal{T}}_0dt\,\dot{W}_{chem}(t)\nonumber\\&=&-\frac{1}{\mathcal{T}}\int^{\mathcal{T}}_0dt\,\sum_m\dot{\mu}(t)n_mp_m(t)\nonumber\\&=&\frac{T_0F_{\mu}}{\mathcal{T}}\int^{\mathcal{T}}_0dt\langle \delta{\boldsymbol{\gamma}}_{\mu}\vert \boldsymbol{p}(t) \rangle\dot{g}_{\mu}(t)\nonumber\\
&=&T_0F_\mu J_\mu.\label{chemwork}\end{eqnarray}

The heat flow to the system can be determined by noting that there is, in the time-periodic steady state, per period no net energy flux into the system. Hence the first law of thermodynamics dictates that:
\begin{equation}\dot{\overline{W}}+\dot{\overline{W}}_{chem}+\dot{\overline{Q}}=0,\label{fl}\end{equation}
leading to:
\begin{eqnarray}
\dot{\overline{Q}}&=&\frac{1}{\mathcal{T}}\int^{\mathcal{T}}_0dt\,\dot{Q}(t)\nonumber\\&=&\frac{1}{\mathcal{T}}\sum_m\int^{\mathcal{T}}_0dt\,\left(\epsilon_m(t)-\mu(t)n_m\right)\dot{p}_m(t).\label{qdef}
\end{eqnarray}

The entropy production can on the one hand be defined in terms of the heat flow, and on the other hand in terms of the thermodynamic fluxes and forces \cite{callen2006thermodynamics}:\begin{equation}
\dot{\overline{S}}=-\frac{1}{\mathcal{T}}\int^{\mathcal{T}}_0dt\,\frac{\dot{{Q}}(t)}{T(t)} \equiv F_{\mu}J_\mu+F_{T}J_T+F_{\epsilon}J_{\epsilon}.\label{Sdef1}
\end{equation}
Using the definitions of the mechanical and chemical work fluxes introduced earlier, we deduce the following expression for the energy flux associated with the modulation of the temperature:
\begin{eqnarray}J_T&=&\frac{\dot{\overline{S}}-F_\mu J_\mu- F_{\epsilon}J_{\epsilon}}{F_T}\nonumber\\&=&-\frac{1}{\mathcal{T}}\int^{\mathcal{T}}_0dt\,g_T(t)\dot{{Q}}(t)\nonumber\\&=&\frac{1}{\mathcal{T}}\int^{\mathcal{T}}_0dt\,\sum_m(\epsilon_m(t)-\mu(t)n_m)p_m(t)\dot{g}_T(t)\nonumber\\&\approx&\frac{1}{\mathcal{T}}\int^{\mathcal{T}}_0dt\langle \delta{\boldsymbol{\gamma}}_{T}\vert \boldsymbol{p}(t)\rangle\dot{g}_{T}(t).\end{eqnarray}
In the last step, we apply the linear approximation in the thermodynamic forces by setting $\epsilon_m(t)-\mu(t) n_m\approx \epsilon_{0,m}-\mu_{0,m} n_m$.

Summarizing, we have identified the following thermodynamic fluxes:
\begin{eqnarray}
J_{\epsilon}&=&\frac{1}{\mathcal{T}}\int^{\mathcal{T}}_0dt\langle \delta{\boldsymbol{\gamma}}_{\epsilon}\vert \boldsymbol{p}(t) \rangle\dot{g}_{\epsilon}(t),\label{mechflux} \\
J_\mu&=&\frac{1}{\mathcal{T}}\int^{\mathcal{T}}_0dt\langle \delta{\boldsymbol{\gamma}}_{\mu}\vert \boldsymbol{p}(t) \rangle\dot{g}_{\mu}(t), \label{chemflux}\\
J_T&=&\frac{1}{\mathcal{T}}\int^{\mathcal{T}}_0dt\langle \delta{\boldsymbol{\gamma}}_{T}\vert \boldsymbol{p}(t) \rangle\dot{g}_{T}(t) \label{heatflux},
\end{eqnarray}
up to linear order. They appear in combination with the thermodynamic forces $F_{\epsilon}$, $F_{\mu}$ and $F_{T}$ in the usual expression for the entropy production, Eq.~(\ref{Sdef1}).

\section{General theory \label{Gt}}
We now derive the linear stochastic thermodynamics for a system in contact with a single reservoir. 
The interest of this derivation is that it is quite general, and that we can obtain surprisingly explicit results, even though
 we need not specify neither the type of reservoir nor the set of variables.
We use the abstract notation ${\alpha}$ to designate the set of periodically driven variables :
\begin{equation}{\alpha}(t)={\alpha}_0+F_\alpha {\alpha}_1 g_\alpha(t)\;\;;\;\; g_\alpha(t+\mathcal{T})=g_\alpha(t),\;\; \forall \alpha,\end{equation}
with $\mathcal{T}$ the period of the driving and ${\alpha}_0$ the reference values of the parameters. 
${\alpha}$ can be a scalar, such as $1/T_0$ and $\mu/T_0$ or vectorial, such as $\boldsymbol{\epsilon}/T_0$. The amplitudes $F_\alpha$ are the corresponding generalised thermodynamic forces. 
The reservoir is defined by the fact that, in the limit of a slow perturbation (or fast relaxation of the system), the probability distribution 
reduces to the instantaneous adiabatic distribution $\boldsymbol{p}^{\ad}(t)$ (cf.~Eq.~(\ref{dbdef})), which is the equilibrium state in contact with the reservoir for the instantaneous value of 
the parameters. 
This distribution can be expanded in terms of the thermodynamic forces as follows:
\begin{equation}
p^{\ad}_m(t)\approx p^{\eq}_m-\sum_\alpha F_\alpha\delta\gamma_{\alpha,m}g_\alpha(t).\label{pdbgen}\end{equation}
Here, $\boldsymbol{p}^{\eq}$ is the equilibrium distribution in the absence of any driving. 
Eq.~(\ref{pdbgen}) allows to identify 
the quantities $\delta \boldsymbol{\gamma}_\alpha$, which will play a central role in the definition of the fluxes. Furthermore, we see that, as a result of normalization, $\sum_m \delta \gamma_{\alpha,m}=0$, again in agreement with the result from the previous section.

We note in passing that there is some freedom in the definition of the forces. Indeed, the above expressions are invariant under the transformation
\begin{equation}F_\alpha\rightarrow cF_\alpha,\;\;\;{\alpha}_1g_\alpha(t)\rightarrow\frac{{\alpha}_1g_\alpha(t)}{c},\end{equation}
for any $c\neq 0$. We shall fix the choice by assuming that:
\begin{equation}\frac{1}{\mathcal{T}}\int^{\mathcal{T}}_0dt\,\left({\alpha}_1g_\alpha(t)\right)^2=1.\label{ForceNor}\end{equation}
Other conditions are possible, such as $\max_tg_\alpha(t)=1$ \cite{brandner2015thermodynamics}, but the above definition is more convenient when considering a Fourier analysis (see section \ref{fouran}).

In order to proceed, and to evaluate entropy production and fluxes, we need to derive further information about the probability distribution $\boldsymbol{p}(t)$ in its time-periodic steady state:
$\boldsymbol{p}(t+\mathcal{T})=\boldsymbol{p}(t)$. To do so,
we formally expand the transition matrix and the probability distributions up to first order in the thermodynamic forces:
\begin{eqnarray}
\boldsymbol{W}(t)&\approx&\boldsymbol{W}^{\eq}+\sum_\alpha F_\alpha \boldsymbol{W}^{\alpha}(t),\label{Wdef}\\
\boldsymbol{p}\left(t\right)&\approx&\boldsymbol{p}^{\eq}+\sum_\alpha F_\alpha\boldsymbol{p}^{\alpha}(t),\label{Pdef}\\
\boldsymbol{p}^{\ad}(t)&\approx&\boldsymbol{p}^{\eq}+\sum_\alpha F_\alpha\boldsymbol{p}^{\ad,\alpha}(t),
\end{eqnarray}
where $\boldsymbol{W}^{\eq}$ and $\boldsymbol{p}^{\eq}$ are the equilibrium transition matrix and probability distribution if all driven variables are equal to their reference values ${\alpha}_0$. The superscript $\alpha$ refers to the partial derivative of the quantity with respect to a specific force $F_\alpha$, 
and subsequently setting all forces equal to $0$.
Substitution into the master equation, Eq.~(\ref{MEq}) gives:
\begin{equation}\dot{\boldsymbol{p}}^{\alpha}(t)=\boldsymbol{W}^{\eq}\boldsymbol{p}^\alpha(t)+\boldsymbol{W}^{\alpha}(t)\boldsymbol{p}^{\eq}.\label{pdefex}\end{equation}
Expanding the equation obeyed by the adiabatic probability distribution, Eq.~(\ref{dbdef}), in the thermodynamic forces gives:
\begin{equation}\boldsymbol{W}^{\alpha}(t)\boldsymbol{p}^{\eq}+\boldsymbol{W}^{\eq}\boldsymbol{p}^{\ad,\alpha}(t)=0.\label{dbdefex}\end{equation}
Substitution of Eq.~(\ref{dbdefex}) into the solution of Eq.~(\ref{pdefex}) for $\boldsymbol{p}^\alpha(t)$, finally leads to \cite{risken1984fokker}:
\begin{equation}
\boldsymbol{p}\left(t\right)\approx\boldsymbol{p}^{\ad}\left(t\right)-\int^{\infty}_0d\tau e^{\boldsymbol{W}^{\eq}\tau} \dot{\boldsymbol{p}}^{\ad}(t-\tau).\label{ppdb}
\end{equation}
We conclude that the solution of the master equation can be written, up to first order in the thermodynamic forces, purely in terms of the reference transition matrix and the adiabatic distribution.

\subsection{Entropy production}
Since we have not specified the modulated variables ${\alpha}$, we need, in order to obtain the entropy production, not start from the heat flux as in Eq.~(\ref{Sdef1}), but rather from the general formula for entropy production of systems described by a master equation \cite{jiu1984stability,schnakenberg1976network}. By averaging over one period, one has:
\begin{equation}\dot{\overline{S}}=\frac{1}{\mathcal{T}}\sum_{n,m}\int^{\mathcal{T}}_0dt\,W_{nm}(t)p_m(t)\ln\left(\frac{W_{nm}(t)p_m(t)}{W_{mn}(t)p_n(t)}\right).\label{Schnaken}\end{equation}
The resulting explicit expression for entropy production, cf.~Eq.~(\ref{entrodet}) below, is obtained in \ref{appep} starting from this formula.
Here we reproduce this result via a shortcut by noting that
the entropy production can be split into an adiabatic and a non-adiabatic part \cite{esposito2010det,esposito2010three}. In general, the adiabatic term corresponds to a contribution of a reference (time-independent) steady state, while the non-adiabatic entropy production refers to a "relaxational" contribution. In our case of a single reservoir, the reference state is always an equilibrium (detailed-balance) distribution, and therefore the adiabatic entropy is equal to zero. We conclude that the entropy production has to be purely non-adiabatic, hence given by the following expression \cite{esposito2010det,esposito2010three}:
\begin{equation}\dot{\overline{S}}=-\frac{1}{\mathcal{T}}\int^{\mathcal{T}}_0dt\,\sum_m\dot{p}_m(t)\ln\left(\frac{p_m(t)}{p^{\ad}_m(t)}\right).\label{nadent}\end{equation}
Up to linear order, this can be rewritten, cf.~Eq.~(\ref{pdbgen}):
\begin{eqnarray}
\dot{\overline{S}}&\approx&-\frac{1}{\mathcal{T}}\int^{\mathcal{T}}_0dt\,\left\langle\boldsymbol{p}(t)|\dot{\boldsymbol{p}}^{\ad}(t)\right\rangle\nonumber\\
&\approx&\sum_{\alpha}\frac{F_\alpha}{\mathcal{T}}\int^{\mathcal{T}}_0dt\,\langle \delta \boldsymbol{\gamma}_{\alpha}\vert \boldsymbol{p}(t)\rangle\dot{g}_\alpha(t).\label{stss}\label{SGendef}\end{eqnarray}
Filling in the expression for the periodic steady state distribution, Eq.~(\ref{ppdb}), one obtains, after a partial integration and in combination with Eq.~(\ref{pdbgen}), the following explicit expression (in agreement with the direct evaluation from \ref{appep}):
\begin{eqnarray}\dot{\overline{S}}&=&-\sum_{\alpha,\beta}\frac{F_\alpha F_\beta}{\mathcal{T}}\int^{\mathcal{T}}_0dt\int^{\infty}_0d\tau\left\langle\delta\boldsymbol{\gamma}_\alpha|\exp\left(\boldsymbol{W}^{\eq}\tau\right)\delta \boldsymbol{\gamma}_\beta\right\rangle g_\alpha(t)\ddot{g}_\beta(t-\tau)\nonumber\\&=&\sum_{\alpha,\beta}\int^{\infty}_0d\tau\,\langle\langle F_\alpha \dot{g}_\alpha(0)\delta\boldsymbol{\gamma}_\alpha;F_\beta\dot{g}_\beta(-\tau)\delta\boldsymbol{\gamma}_\beta\rangle\rangle,\label{SL}\label{entrodet}
\end{eqnarray}
where we introduced the generalized equilibrium correlation function \cite{brandner2015thermodynamics}:
\begin{equation}\langle\langle \boldsymbol{f}_1(t_1);\boldsymbol{f}_2(t_2)\rangle\rangle=\frac{1}{\mathcal{T}}\int^{\mathcal{T}}_0dt\,\left\langle \boldsymbol{f}_1(t_1+t)|e^{\boldsymbol{W}^{\eq}\left|t_1-t_2\right|}\boldsymbol{f}_2(t_2+t)\right\rangle.\end{equation}


\subsection{Thermodynamic Fluxes}
We next turn to the thermodynamic fluxes. The purpose is to write the entropy production in its standard bi-linear thermodynamic form:
\begin{equation}\label{epfj}
\dot{\overline{S}}=\sum_\alpha F_\alpha J_\alpha=\boldsymbol{F}\boldsymbol{J}.
\end{equation}
The fluxes can be identified by comparison with Eq.~(\ref{SGendef}):
\begin{equation}
J_\alpha=\frac{1}{\mathcal{T}}\int^{\mathcal{T}}_0dt\,\langle \delta \boldsymbol{\gamma}_{\alpha}\vert \boldsymbol{p}(t)\rangle\dot{g}_{\alpha}(t).
\end{equation}
These expressions are in agreement with the ones for the grand canonical reservoir, cf.~Eqs.~(\ref{mechflux})-(\ref{heatflux}), and with previous results from the literature \cite{brandner2015thermodynamics,proesmans2015onsager}. Furthermore, as we proceed to show next, the resulting Onsager matrix has the required symmetry properties.

\subsection{Onsager matrix}
The thermodynamic fluxes are linked, up to first order, to the thermodynamic forces via the so-called Onsager matrix $\boldsymbol{L}$:
\begin{equation}\boldsymbol{J}=\boldsymbol{L}\boldsymbol{F}.\label{JOns}\end{equation}
Before turning to the explicit expressions for the Onsager coefficients, we first split the thermodynamic fluxes in two parts:
\begin{equation}\boldsymbol{J}=\boldsymbol{J}^{\ad}+\boldsymbol{J}^{\nad}.\end{equation}
The first term is the adiabatic flux, i.e., the flux in the limit when the distribution is all the time in the adiabatic state:
\begin{equation}
{J}^{\ad}_\alpha=\frac{1}{\mathcal{T}}\int^{\mathcal{T}}_0dt\langle \delta \boldsymbol{\gamma}_{\alpha}\vert \boldsymbol{p}^{\ad}(t)\rangle\dot{g}_{\alpha}(t).
\end{equation}
The second term, i.e.~the non-adiabatic flux, represents the correction due the deviation from adiabaticity:
\begin{equation}
{J}^{\nad}_\alpha=\frac{1}{\mathcal{T}}\int^{\mathcal{T}}_0dt\langle \delta \boldsymbol{\gamma}_{\alpha}\vert \boldsymbol{p}(t)-\boldsymbol{p}^{\ad}(t)\rangle\dot{g}_{\alpha}(t).
\end{equation}
Similarly, we can split the Onsager matrix into an adiabatic contribution $\boldsymbol{L}^{\ad}$ and a non-adiabatic one $\boldsymbol{L}^{\nad}$:
\begin{eqnarray}
\boldsymbol{J}^{\ad}=\boldsymbol{L}^{\ad}\boldsymbol{F}\;\;;\;\;\boldsymbol{J}^{\nad}=\boldsymbol{L}^{\nad}\boldsymbol{F},\;\;\\\boldsymbol{L}=\boldsymbol{L}^{\ad}+\boldsymbol{L}^{\nad}.
\end{eqnarray}
For the adiabatic Onsager coefficient one finds:
\begin{equation}
L^{\ad}_{\alpha\beta}=\frac{1}{\mathcal{T}}\left\langle \delta \boldsymbol{\gamma}_\alpha|\delta\boldsymbol{\gamma}_\beta\right\rangle\int^{\mathcal{T}}_0dt\,g_\alpha(t)\dot{g}_\beta (t).\label{Lad}
\end{equation}
Clearly, this matrix is purely anti-symmetric: $L^{\ad}_{\alpha\beta}=-L^{\ad}_{\beta\alpha}$. The non-adiabatic Onsager matrix on the other hand can be written as:
\begin{eqnarray}
L^{\nad}_{\alpha\beta}&=&-\frac{1}{\mathcal{T}}\int^{\mathcal{T}}_0dt\int^{\infty}_0d\tau\,\left\langle \delta \boldsymbol{\gamma}_\alpha|e^{\boldsymbol{W}^{\eq}\tau}\delta \boldsymbol{\gamma}_\beta\right\rangle g_\alpha(t) \ddot{g}_\beta(t-\tau)\nonumber\\&=&\int^{\infty}_0d\tau\,\langle\langle \dot{g}_\alpha(0)\delta\boldsymbol{\gamma}_\alpha;\dot{g}_\beta(-\tau)\delta\boldsymbol{\gamma}_\beta\rangle\rangle.\label{Lnad}
\end{eqnarray}
We proceed to show that the Onsager coefficients obey a generalized Onsager-Casimir relation \cite{casimir1945onsager,brandner2015thermodynamics,proesmans2015onsager}
\begin{eqnarray}
L_{\alpha\beta}=\tilde{L}_{\beta\alpha},
\label{OnsCas}\end{eqnarray}
where $\tilde{\boldsymbol{L}}$ corresponds to the Onsager matrix for the time-reversed perturbation $\tilde{g}_\gamma(t)=g_\gamma(-t)$.
For time-symmetric driving, for $\tilde{g}_\gamma(t)=g_\gamma(t)$, one recovers the "usual" Onsager symmetry relation: $L_{\alpha\beta}=L_{\beta\alpha}$ \cite{onsager1931reciprocal,onsager1931reciprocal2}.
The proof of Eq.~(\ref{OnsCas}) goes as follows. One can write, starting from Eq.~(\ref{Lnad}), that:
\begin{eqnarray}
L^{\nad}_{\alpha\beta}\left[\tilde{g}_\gamma(t)\right]&&=-\frac{1}{\mathcal{T}}\int^{\mathcal{T}}_0dt\int^{\infty}_0d\tau \left\langle \delta \boldsymbol{\gamma}_\alpha|e^{\boldsymbol{W}^{\eq}\tau}\delta \boldsymbol{\gamma}_\beta\right\rangle g_\alpha(-t) \ddot{g}_\beta(-t+\tau)\nonumber\\&&=-\frac{1}{\mathcal{T}}\int^{\mathcal{T}}_0dt'\int^{\infty}_0d\tau \left\langle \delta \boldsymbol{\gamma}_\alpha|e^{\boldsymbol{W}^{\eq}\tau}\delta \boldsymbol{\gamma}_\beta\right\rangle g_\alpha(t'-\tau) \ddot{g}_\beta(t')\nonumber\\&&=L^{\nad}_{\beta\alpha}\left[g_\gamma(t)\right],
\end{eqnarray}
where we have made a change of variables: $t'=-t+\tau$, and shifted the integration bounds of the $t'$ integral. This does not modify the integral, as the $g$-functions are periodic. An analogous argument holds for the adiabatic part of the Onsager matrix, and therefore for the total Onsager matrix. More detailed symmetry properties will be revealed in the next section, when focusing on the "detailed" Onsager matrix that appears when expanding the perturbation in terms of its Fourier coefficients. 

Returning to the entropy production, one finds by combination of Eqs.~(\ref{epfj}) and (\ref{JOns}):
\begin{equation}\dot{\boldsymbol{S}}=\boldsymbol{F}\boldsymbol{L}\boldsymbol{F},\end{equation}
in agreement with the previous expression, Eq.~(\ref{SL}). It is clear that only the symmetric part of the Onsager matrix contributes to the entropy production. The adiabatic fluxes and the adiabatic Onsager matrix do not give rise to entropy production, as was already anticipated in our discussion about the adiabatic and non-adiabatic entropy production.

\section{Fourier analysis}\label{fouran}
To get further insight, it is revealing to perform the Fourier analysis of the signals. Since all signals 
have a common period, one can write:
\begin{equation}
g_\alpha(t)=\sum_n a_{(\alpha,n,s)}\sin\left(\frac{2\pi n}{\mathcal{T}}t\right)+a_{(\alpha,n,c)}\cos\left(\frac{2\pi n}{\mathcal{T}}t\right).
\end{equation}
The normalisation of the driving function $g_\alpha(t)$, Eq.~(\ref{ForceNor}), now dictates that:
\begin{equation}\sum_n a_{(\alpha,n,s)}^2+a_{(\alpha,n,c)}^2=\frac{2}{({\alpha}_1)^2}.\end{equation}
The thermodynamic fluxes, the entropy production and the Onsager coefficients can thus be expanded in terms of the Fourier components $a_{(\alpha,n,\sigma)}$. A direct calculation gives:
\begin{eqnarray}
J_\alpha
&=&\sum_{\beta,n}\sum_{\sigma,\rho=s,c}-2\delta_{\sigma,\rho}F_\beta\pi^2n^2a_{(\alpha,n,\sigma)}a_{(\beta,n,\rho)}\left\langle\delta \boldsymbol{\gamma}_\alpha \bigg|\boldsymbol{W}^{\eq}\left(4\pi^2n^2+\left(\boldsymbol{W}^{\eq}\mathcal{T}\right)^2\right)^{-1}\delta\boldsymbol{\gamma}_\beta\right\rangle \nonumber\\
&&+F_\beta (-1)^{\delta_{\sigma,s}}(1-\delta_{\sigma,\rho})a_{(\alpha,n,\sigma)}a_{(\beta,n,\rho)}\pi n\mathcal{T}\nonumber\\&&\qquad\quad\quad\qquad\quad\qquad\times\left\langle \delta \boldsymbol{\gamma}_\alpha\bigg|\left(\boldsymbol{W}^{\eq}\right)^2\left(4\pi^2n^2+\left(\boldsymbol{W}^{\eq}\mathcal{T}\right)^2\right)^{-1}\delta\boldsymbol{\gamma}_\beta\right\rangle,\label{JDet}
\end{eqnarray}
\begin{equation}
\dot{\overline{S}}=\sum_{\alpha,\beta,n,\sigma}-2F_\alpha F_\beta\pi^2n^2a_{(\alpha,n,\sigma)}a_{(\beta,n,\sigma)}\left\langle\delta \boldsymbol{\gamma}_\alpha \bigg|\boldsymbol{W}^{\eq}\left(4\pi^2n^2+\left(\boldsymbol{W}^{\eq}\mathcal{T}\right)^2\right)^{-1}\delta\boldsymbol{\gamma}_\beta\right\rangle,
\end{equation}
\begin{equation}
L_{\alpha\beta}^{\ad}=\frac{1}{\mathcal{T}}\left\langle \delta\boldsymbol{\gamma}_\alpha|\delta \boldsymbol{\gamma}_\beta\right\rangle \sum_{n,\sigma,\rho}(1-\delta_{\sigma,\rho})(-1)^{\delta_{\sigma,s}}\pi na_{(\alpha,n,\sigma)}a_{(\beta,n,\rho)},\label{Ladft}
\end{equation}
\begin{eqnarray}
L_{\alpha\beta}^{\nad}&=&\sum_{n,\sigma,\rho}-2\delta_{\sigma,\rho}\pi^2n^2a_{(\alpha,n,\sigma)}a_{(\beta,n,\rho)}\left\langle\delta \boldsymbol{\gamma}_\alpha \bigg|\boldsymbol{W}^{\eq}\left(4\pi^2n^2+\left(\boldsymbol{W}^{\eq}\mathcal{T}\right)^2\right)^{-1}\delta\boldsymbol{\gamma}_\beta\right\rangle\nonumber\\&&
-(1-\delta_{\sigma,\rho})(-1)^{\delta_{\sigma,s}}a_{(\alpha,n,\sigma)}a_{(\beta,n,\rho)}\frac{4\pi^3n^3}{\mathcal{T}}\left\langle\delta \boldsymbol{\gamma}_\alpha \bigg|\left(4\pi^2n^2+\left(\boldsymbol{W}^{\eq}\mathcal{T}\right)^2\right)^{-1}\delta\boldsymbol{\gamma}_\beta\right\rangle.\nonumber\\
\end{eqnarray}
These results allow to make both general and detailed statements about the thermodynamic properties of the system. We give an illustration in the next section by optimizing the power of a quantum dot and make a general statement about the zero entropy production regime in \ref{zep}. The Onsager-Casimir relations also acquire a more profound interpretation, linking it to the symmetry properties of the geometric functions. As the sine function is time anti-symmetric and the cosine function is time symmetric, one has:
\begin{equation}\tilde{a}_{(\alpha,n,\sigma)}=(-1)^{\delta_{\sigma,s}}a_{(\alpha,n,\sigma)},\end{equation}
where the tilde stands for time-inversion.
Using these symmetries in the expressions for the adiabatic and the non-adiabatic Onsager coefficients immediately leads to the generalized Onsager-Casimir relations, Eq.~(\ref{OnsCas}).

We finally discuss the limits of fast ($\mathcal{T}\rightarrow 0$) and slow ($\mathcal{T}\rightarrow \infty$) driving. For fast driving, the adiabatic Onsager matrix is cancelled by the anti-symmetric term in the non-adiabatic Onsager matrix. Hence the Onsager matrix reduces to the symmetric part of the non-adiabatic contribution. In particular, all expressions become invariant under time-inversion. In the limit of slow driving, both the adiabatic and non-adiabatic coefficients vanish but at a different rate: $L^{\ad}_{\alpha\beta}\sim O\left(1/\mathcal{T} \right)$ whereas $L^{\nad}_{\alpha\beta}\sim O\left(1/\mathcal{T}^2 \right)$. This result can be understood intuitively by noting that for $\mathcal{T}\rightarrow \infty$ the functions $g_{\alpha}(t)$ are stretched and as a result their time derivatives are reduced. Since $L^{\nad}_{\alpha\beta}$ depends on the second order time derivative, as opposed to $L^{\ad}_{\alpha\beta}$ which depends only on the first order derivative, it vanishes faster than $L^{\ad}_{\alpha\beta}$. The Onsager matrix then obviously reduces to its anti-symmetric adiabatic form. An additional interesting observation for this limit is that the entropy production goes to zero faster than the fluxes, hence the system behaves as an adiabatic pump \cite{cohen2003quantum,sinitsyn2009stochastic}.

\section{Example: modulated two-level systems \label{exa}}
\begin{figure}\begin{centering}
\includegraphics[scale=0.4]{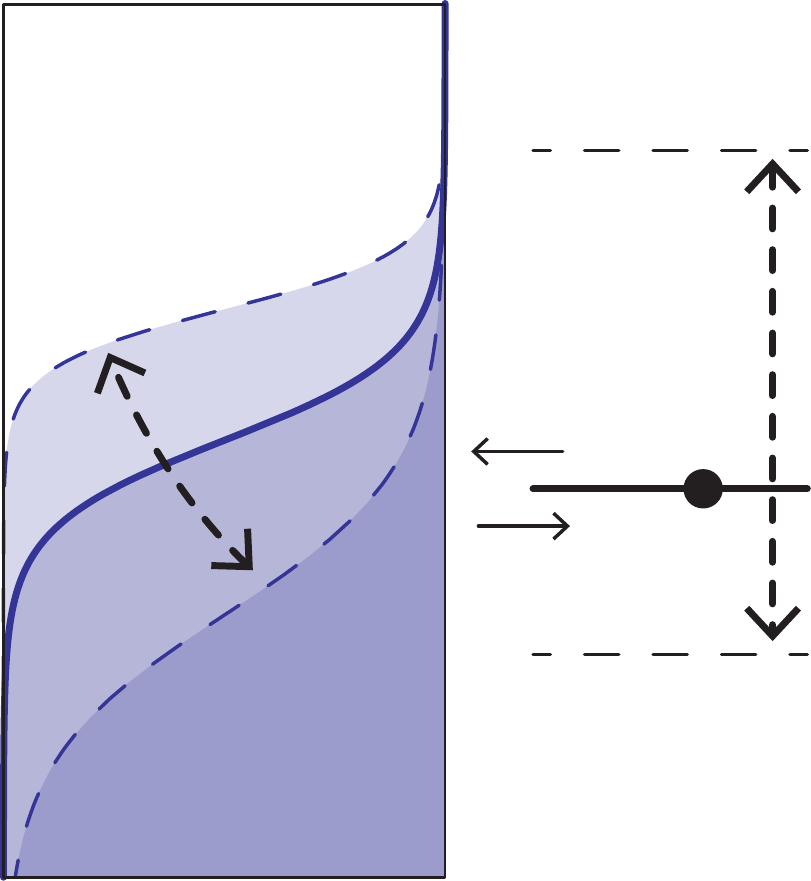}
\caption{(Color online) Schematic representation of the periodically driven quantum dot.\label{QD}}
\end{centering}\end{figure}
To illustrate our results, we turn to a specific example: a system consisting of a quantum dot with a single active energy level, which is periodically modulated $\epsilon(t)/T_0=\epsilon_0/T_0+F_\epsilon g_\epsilon(t)$. The quantum dot is either empty, with probability $p_0(t)$, or filled by a single electron with probability $p_1(t)=1-p_0(t)$. It is in contact with an electron reservoir at inverse temperature $1/T(t)=1/T_0+F_Tg_T(t)$ and chemical potential $\mu(t)/T_0=\mu_0/T_0+F_\mu g_\mu(t)$, cf.~Fig.~\ref{QD}. The rates of transition into and out of the quantum dot, $W_{10}$ and $W_{01}$  respectively, reproduce Fermi-Dirac statistics:
\begin{eqnarray}
W_{10}=\Gamma f(x(t))=\Gamma\left[1+\exp(x(t))\right]^{-1} \;\; ;\;\;W_{01}=\Gamma (1-f(x(t))\;\; ;\nonumber\\\qquad\quad\qquad\qquad\qquad\qquad\qquad\qquad\qquad x(t)=\left(\epsilon(t)-\mu(t)\right)/T(t).
\end{eqnarray}
$\Gamma$ quantifies the coupling strength between quantum dot and reservoir.
Introducing
\begin{eqnarray}g(t)&=&F_T(\epsilon_0-\mu_0){g}_T(t)+F_\epsilon{g}_\epsilon(t)-F_\mu {g}_\mu (t)\\&=&\sum_n a_{(n,s)}\sin\left(\frac{2\pi n}{\mathcal{T}}t\right)+a_{(n,c)}\cos\left(\frac{2\pi n}{\mathcal{T}}t\right),\nonumber\end{eqnarray}
with
\begin{equation}a_{(n,\sigma)}=F_T(\epsilon_0-\mu_0) a_{(T,n,\sigma)}+F_\epsilon a_{(\epsilon,n,\sigma)}-F_\mu a_{(\mu,n,\sigma)},\end{equation}
the expressions for the work and heat fluxes become:
\begin{eqnarray}
\dot{\overline{W}}&=&\sum_n \frac{F_{\epsilon}T_0p^{\eq}_0p^{\eq}_1\chi\pi n}{\mathcal{T}\left(4\pi^2n^2+\chi^2\right)}
\begin{array}{l}
\bigg[2\pi n\left(a_{(\epsilon,n,s)}a_{(n,s)}+a_{(\epsilon,n,c)}a_{(n,c)}\right)\\
\;\;\;\;-\chi\left(a_{(\epsilon,n,s)}a_{(n,c)}-a_{(\epsilon,n,c)}a_{(n,s)}\right)\bigg]
\end{array}
,\label{QDW}
\end{eqnarray}
\begin{eqnarray}
\dot{\overline{W}}_{chem}&=&-\sum_n\frac{{F_{\mu}}T_0p^{\eq}_0p^{\eq}_1\chi\pi n}{\mathcal{T}\left(4\pi^2n^2+\chi^2\right)}
\begin{array}{l}
\bigg[2\pi n\left(a_{(\mu,n,s)}a_{(n,s)}+a_{(\mu,n,c)}a_{(n,c)}\right)\\
\;\;\;\;-\chi\left(a_{(\mu,n,s)}a_{(n,c)}-a_{(\mu,n,c)}a_{(n,s)}\right)\bigg]
\end{array},
\end{eqnarray}
\begin{eqnarray}
\dot{\overline{Q}}&=&\sum_n\frac{T_0p^{\eq}_0p^{\eq}_1\chi\pi n}{\mathcal{T}\left(4\pi^2n^2+\chi^2\right)}
\begin{array}{l}
\bigg[
-2\pi n(a_{(n,s)}^2+a_{(n,c)}^2) +(\epsilon_0-\mu_0)F_T\\
\;\;\;\;\;\times \left[2\pi n\left(a_{(T,n,s)}a_{(n,s)}+a_{(T,n,c)}a_{(n,c)}\right)\right.\\
\;\;\;\;\;\;\;\;\;\;-\chi \left.\left(a_{(T,n,s)}a_{(n,c)}-a_{(T,n,c)}a_{(n,s)}\right)\right]\bigg]
\end{array},
\end{eqnarray}
where we introduced $\chi=\Gamma\mathcal{T}$, a dimensionless number which is an inverse measure of the distance between the exact and the adiabatic probability distribution. Notice the analogy between the expressions, except for an extra term in the heat flux. This term guarantees that, in the absence of temperature driving, the heat flux to the system is always negative, i.e., work is dissipated as heat, in agreement with 'classical' thermodynamics.

We now consider the situation of a heat engine and set $F_\mu=0$, hence the chemical work flux is zero $\dot{\overline{W}}_{chem}=0$. Furthermore, we assume that the thermodynamic forces $F_T$ and $F_\epsilon$ are small so that all expressions can be linearised. Such a setup allows for a positive power output in the following way. First the system is cooled while the energy is increased. The quantum dot is then most likely empty so that no work is required to lift the energy level. Next, the temperature of the reservoir is raised. As a result the probability for the quantum dot to be occupied increases. Lowering the energy level, while keeping the temperature high, thus results in work delivery on the work reservoir. The net result of the complete cycle is a positive power output. The efficiency of this procedure can be quantified by considering the entropy production during the total cycle, $\dot{\overline{S}}=F_{T}J_T+F_{\epsilon}J_{\epsilon}$. A positive dissipative contribution $F_{T}J_T $ is compatible with a negative contribution for $F_\epsilon J_\epsilon$, with the following ratio quantifying the corresponding efficiency: 
\begin{equation}
\eta=-\frac{F_\epsilon J_\epsilon}{F_TJ_T} \leq 1.\label{etamaxp}
\end{equation}
The upper bound is reached for a reversible operation $\dot{\overline{S}}=0$.

For the purpose of illustration, we limit ourselves to driving at a single frequency, more precisely:
\begin{eqnarray}
g_{T}(t)&=&\sqrt{2}\sin\left(\frac{2\pi}{\mathcal{T}}t\right);\\
g_{\epsilon}(t)&=&\sqrt{2}\sin\left(\frac{2\pi}{\mathcal{T}}t+\phi\right)
\end{eqnarray}
with $\epsilon_0>\mu_0$ and $\phi$ the phase-difference between the driving of the energy level and the temperature. The mechanical work reduces to:
\begin{eqnarray}
\dot{\overline{W}}&=&\frac{{2\pi\chi p^{\eq}_0p^{\eq}_1 F_{\epsilon}}T_0}{\mathcal{T}\left(4\pi^2+\chi^2\right)}\nonumber\\&&\Big[2\pi F_\epsilon+ F_T\chi(\epsilon_0-\mu_0)\sin(\phi)+2\pi F_T(\epsilon_0-\mu_0)\cos(\phi)\Big].\label{QDW2}
\end{eqnarray}
The entropy production is given by:
\begin{eqnarray}
\dot{\overline{S}}&=&\frac{4\pi^2\chi p^{\eq}_0p^{\eq}_1}{\mathcal{T}\left(4\pi^2+\chi^2\right)}\Big[F_\epsilon^2+2F_\epsilon F_T(\epsilon_0-\mu_0)\cos(\phi)+F_T^2(\epsilon_0-\mu_0)^2\Big].\label{SExa}\end{eqnarray}

\begin{figure}\begin{centering}
\includegraphics[scale=0.38]{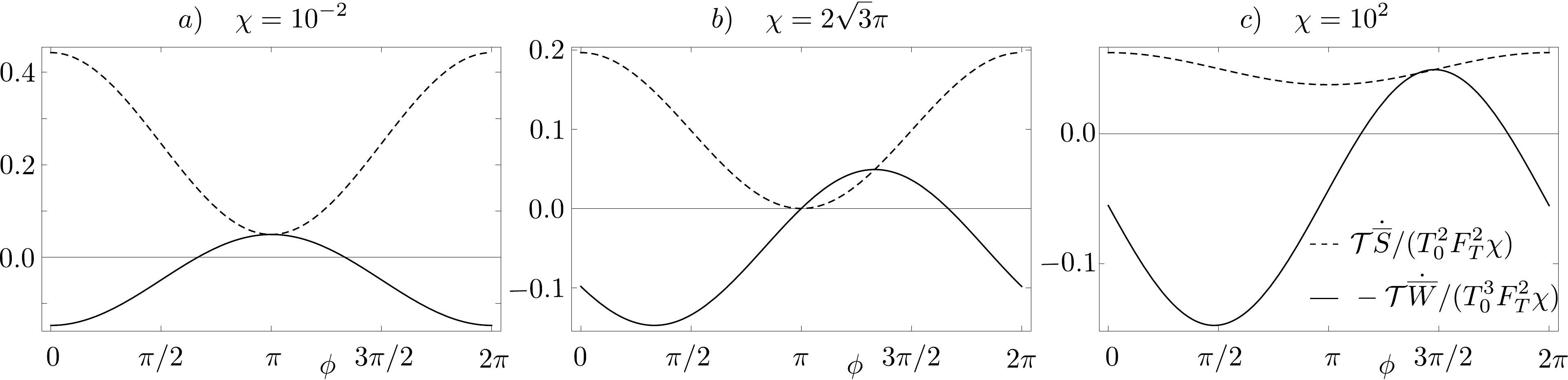}
\caption{(Color online) Entropy production $\mathcal{T}\dot{\overline{S}}/(T_0^2F_T^2\chi)$ (dashed lines) and work output $-\mathcal{T}\dot{\overline{W}}/(T_0^3F_T^2\chi)$ (full lines) as a function of $\phi$, for a) $\chi=10^{-2}$, b) $\chi=2\sqrt{3}\pi$ and c) $\chi=10^2$. Parameter values are $(\epsilon_0-\mu_0)/T_0=1$ and $F_{\epsilon}$ determined by Eq.~(\ref{maxPowForce}). The linear approximation is assumed to be valid ($T_0F_T\ll 1$). Panel (b) provides an example of zero entropy production, reached for $\phi=\pi$ and $\chi=2\sqrt{3}\pi$. Furthermore $\dot{\overline{W}}_{Max}$ is independent of $\chi/\mathcal{T}$. In the limit $\chi\rightarrow 0$ (panel (a)), the maximum power is reached for $\phi=\pi$ whereas in the opposite limit $\chi\rightarrow \infty$ (panel (c)), this maximum appears for $\phi=3\pi/2$. In addition, the efficiency at maximum power satisfies $-\dot{\overline{W}}_{Max}=T_0\dot{\overline{S}}$, as the two curves touch or cross in this point.
\label{MaxPow}}
\end{centering}\end{figure}

Several interesting situations can now be considered. We first identify the conditions for reversible operation (at least a this linear order in the forces). The entropy production is a nonnegative quadratic expression in $F_{\epsilon}/F_{T}$. 
Hence, zero entropy production with nonvanishing thermodynamic forces is only possible when the discriminant is equal to zero $4(\epsilon_0-\mu_0)^2\sin^2(\phi)=0$, i.e., $\phi=\pi$ or $\phi=0$ (mod $2\pi$) . Under this condition, zero entropy production is realized for a corresponding ratio of forces $F_{\epsilon}/F_{T}=\pm (\epsilon_0-\mu_0)$. An example of such a situation is represented in Fig.~\ref{MaxPow} for $\phi=\pi$ with $F_{\epsilon}/F_{T}=\epsilon_0-\mu_0$. It might come as a surprise that one can reproduce equilibrium conditions in a modulated system. The explanation is simple: this will happen when the modulations are such that the adiabatic solution $ \boldsymbol{p}^{\ad}(t)$ is time-independent, which becomes the de facto new equilibrium state. This condition is satisfied under the above mentioned conditions, with the important proviso that it is only valid up to linear order in the thermodynamic forces. Hence, in the present example, one thus not reproduce a genuine equilibrium state.

Next, we focus on the work protocol that gives maximum power output. More precisely, we keep $F_T$ fixed, and optimise the power Eq.~(\ref{QDW2}) with respect to $F_\epsilon$ and $\phi$. This leads to:
\begin{equation}\label{maxPowForce}
\frac{F_\epsilon}{F_T}=\frac{\sqrt{4\pi^2+\chi^2}}{4\pi}(\epsilon_0-\mu_0)\;\;\;\;;\;\;\;\;\phi=\tan^{-1}\left(\frac{\chi}{2\pi}\right)+\pi.\end{equation}
Note that $\phi$ only depends on $\chi$.
In the adiabatic or strong coupling limit $\chi\rightarrow\infty$, the optimal driving amplitude diverges and the phase difference becomes equal to $3\pi/2$. This means that the speed at which the energy of the quantum dot decreases is maximal when the temperature is at its peak, corresponding to the maximum in the occupation probability of the quantum dot, and the corresponding maximum amount of delivered work.
In the limit of fast driving or weak coupling $\chi\rightarrow 0$ on the other hand, the system cannot relax to its equilibrium state and the phase difference during optimal driving converges to $\pi$. This implies that at the moment the energy level starts dropping, the temperature of the quantum dot has been growing so that its occupation probability is high and the system again delivers work.
The corresponding efficiency at maximum power is found by insertion of Eq.~(\ref{maxPowForce}) into the expression for power, Eq.~(\ref{QDW2}), and entropy production, Eq.~(\ref{SExa}). The following simple relationship holds:
\begin{equation}
-\dot{\overline{W}}_{\mathrm{Max}}=\frac{\chi p^{\eq}_0p^{\eq}_1 F_T^2T_0 (\epsilon_0-\mu_0)^2}{4\mathcal{T}}=T_0\dot{\overline{S}}.\label{wspow}
\end{equation}
We thus conclude, cf.~Eq.~(\ref{etamaxp}), that the efficiency at maximum power $\eta_{MP}$ is exactly equal to half of the reversible efficiency:
$\eta_{MP}={1}/{2}$.
This result is reminiscent of the similar general result obtained from 'classical' linear thermodynamics under time-symmetric driving \cite{van2005thermodynamic,van2012efficiency,Cleuren2015Universality}.

Finally, as an illustration of the various points from the above discussion, we reproduce in Fig.~\ref{MaxPow} the scaled entropy production $\mathcal{T}\dot{\overline{S}}/(T_0^2F_T^2\chi)$ and power $-\mathcal{T}\dot{\overline{W}}/(T_0^3F_T^2\chi)$ as a function of $\phi$, for different values of $\chi$. 

\section{Discussion and Outlook \label{disc}}
We have further developed the theory of linear stochastic thermodynamics, initiated in \cite{brandner2015thermodynamics,proesmans2015onsager}, by considering periodically driven systems in contact with a single reservoir. Appropriate thermodynamic forces and fluxes are identified, starting from the entropy production for a Markov process. Onsager coefficients are evaluated, the Onsager-Casimir relations are verified, and explicit expressions are given for an expansion in terms of Fourier components. The main surprise is that this analysis is possible without specifying neither the type of reservoir nor the identity of the modulated variables. This generality allows at the same time to zoom into any specific situation, but also to make very general statements. One such statement is made in the appendix concerning the fact that zero entropy production is incompatible with the existence of nonzero fluxes. Another one, namely the fact that the efficiency at maximum power is bounded by $1/2$, will be discussed in a separate paper. Furthermore, it is rather straightforward to generalize our results to systems with multiple reservoirs. In this case the adiabatic distribution is replaced by a steady-state distribution that in general no longer obeys detailed balance. The entropy production will then no longer consist solely of a non adiabatic contribution. Another promising extension is the application to quantum mechanical systems, as the thermodynamics of periodically driven systems is currently a very active field of research \cite{PhysRevB.89.161306,goldman2014periodically,Cuetara2015Stoch,abanin2015exponentially}. 
We finally mention that several existing experimental set-ups \cite{blickle2006thermodynamics,berut_experimental_2012,jun2014high,koski2014experimental} should allow to verify the predictions of the present theory, and in particular the value and symmetry properties of the generalized Onsager coefficients.

\appendix

\section{Entropy production \label{appep}}
We shall here prove Eq.~(\ref{entrodet}) for the entropy production, starting from Eq.~(\ref{Schnaken}).
Up to linear order, we have, using detailed balance \cite{proesmans2015onsager}:
\begin{eqnarray}\ln\left(\frac{W_{mn}(t)p_n(t)}{W_{nm}(t)p_m(t)}\right)&\approx& \ln\left(\frac{1+\sum_\alpha F_\alpha\frac{W^{\eq}_{mn}p^{\alpha}_{n}(t)-W_{mn}^{\eq}(t)p^{\ad,\alpha}_n}{W^{\eq}_{mn}p^{\eq}_n}}{1+\sum_\alpha F_\alpha\frac{W^{\eq}_{nm}p^{\alpha}_{m}(t)-W^{\eq}_{nm}(t)p^{\ad,\alpha}_m}{W^{\eq}_{nm}p^{\eq}_m}}\right)\nonumber\\&\approx&\sum_{\alpha}F_\alpha \left(\frac{p^{\alpha}_{n}(t)-p^{\ad,\alpha}_{n}(t)}{p^{\eq}_n}-\frac{p^{\alpha}_{m}(t)-p^{\ad,\alpha}_{m}(t)}{p^{\eq}_m}\right),\nonumber\\\end{eqnarray}
where we explicitly used the adiabatic relation.
From this, we can write the entropy production as:
\begin{eqnarray}\dot{\overline{S}}&=&\frac{1}{\mathcal{T}}\sum_{mn}\int^{\mathcal{T}}_0dt\,W_{mn}(t)p_n(t)\ln\left(\frac{W_{mn}(t)p_n(t)}{W_{nm}(t)p_m(t)}\right)\nonumber\\&=&\sum_{\alpha,m,n}\frac{F_\alpha}{\mathcal{T}}\int^{\mathcal{T}}_0dt\,W_{mn}(t)p_n(t)\frac{\left(p^{\alpha}_{n}(t)-p^{\ad,\alpha}_{n}(t)\right)}{p^{\eq}_n}\nonumber\\&&-\sum_{\alpha,m}\frac{F_\alpha}{\mathcal{T}}\int^{\mathcal{T}}_0dt\,\dot{p}_m(t)\frac{\left(p^{\alpha}_{m}(t)-p^{\ad,\alpha}_{m}(t)\right)}{p^{\eq}_m}.\end{eqnarray}
The first term is clearly zero, due to the sum over $m$. The entropy production becomes
\begin{equation}\dot{\overline{S}}=-\sum_{\alpha,m}\frac{F_\alpha}{\mathcal{T}}\int^{\mathcal{T}}_0dt\,\dot{p}_m(t)\left(\frac{p^{\alpha}_{m}(t)-p^{\ad,\alpha}_{m}(t)}{p^{\eq}_m}\right).\end{equation}
This result can also be found directly, by starting from the non-adiabatic entropy production Eq.~(\ref{nadent}). This can be seen as an extra proof for the non-adiabaticity of the entropy production. By evaluating the sum over $\alpha$, one sees that the first term is a total derivative of a periodic function integrated over one period, which becomes zero. We then arrive at:
\begin{eqnarray}\dot{\overline{S}}&=&\sum_{\alpha,m}\frac{F_\alpha}{\mathcal{T}}\int^{\mathcal{T}}_0dt\,\frac{p^{\ad,\alpha}_{m}(t)\dot{p}_m(t)}{p^{\eq}_m}\nonumber\\&=&\sum_{\alpha,\beta,m}\frac{F_\alpha F_\beta}{\mathcal{T}}\int^{\mathcal{T}}_0dt\,\left(\frac{p^{\ad,\alpha}_{m}(t)\dot{p}^{\ad,\beta}_m(t)}{p^{\eq}_m}\right.\nonumber\\&-&\left.\sum_n\int^{\infty}_0d\tau\,\frac{p^{\ad,\alpha}_{m}(t)\exp\left(\boldsymbol{W}^{\eq}\tau\right)_{mn}\ddot{p}^{(db,\beta)}_n(t-\tau)}{p^{\eq}_m}\right),\end{eqnarray}
where we used Eq.~(\ref{ppdb}). In this last equation, the first part equals zero as it is a total derivative of a periodic function. Filling in Eq.~(\ref{pdbgen}), now immediately gives Eq.~(\ref{entrodet}).

\section{Zero-entropy production\label{zep}}
To study the limit of zero entropy production, we will now define 'detailed' thermodynamic fluxes, in analogy with \cite{proesmans2015onsager}:
\begin{equation}J_{(n,s/c),k}=\frac{1}{\mathcal{T}}\int^{\mathcal{T}}_0dt\,\dot{g}_{(n,s/c)}(t)\frac{p_k(t)}{p_k^{\eq}},\end{equation}
with $g_{(n,s)}(t)=\sin(2\pi n t/\mathcal{T})$ and analogous for $c$.
With these definitions, we have:
\begin{equation}J_\alpha=\sum_{n,\sigma,k}a_{(\alpha,n,\sigma)}\delta \gamma_{\alpha,k}J_{(n,\sigma),k}.\end{equation}
Furthermore, we can associate detailed thermodynamic forces to these detailed fluxes:
\begin{equation}
F_{(n,\sigma),k}=\sum_\alpha F_\alpha a_{(\alpha,n,\sigma)}\delta \gamma_{\alpha,k}.
\end{equation}
This result can be used to rewrite the entropy production in terms of the detailed thermodynamic fluxes:
\begin{eqnarray}\dot{\overline{S}}&=&\sum_{k,n,\sigma} F_{(n,\sigma),k} J_{(n,\sigma),k}\nonumber\\&=&-\sum_{n,\sigma=s,c}2\pi^2n^2\nonumber\\&&\left\langle \boldsymbol{F}_{(n,\sigma)} |\boldsymbol{W}^{\eq}\left(4\pi^2n^2+\left(\boldsymbol{W}^{\eq}\mathcal{T}\right)^2\right)^{-1}\boldsymbol{F}_{(n,\sigma)}\right\rangle.\nonumber\\\label{SDetf}\end{eqnarray}
With these definitions, the entropy production can only go to zero if
$\boldsymbol{F}_{(n,\sigma)}\sim \boldsymbol{p}^{\eq}$. 
To proof this statement, we decompose $\boldsymbol{F}_{(n,\sigma)}$ in the right eigenvectors of $\boldsymbol{W}^{\eq}$:
\begin{equation}\boldsymbol{F}_{(n,\sigma)}=\sum_i c_i \boldsymbol{r}_i.\end{equation}
General theory predicts that the eigenvector $\boldsymbol{r}_0$, corresponding to eigenvalue $0$, is given by $\boldsymbol{p}^{\eq}$ and that all other eigenvalues $\lambda_i$ are strictly negative. The entropy production can be rewritten as:
\begin{equation}
\dot{\overline{S}}=-\sum_{n,\sigma,i}2\pi^2n^2c_i^2\frac{\lambda_i}{4\pi^2n^2+\lambda^2_i\mathcal{T}^2}\left\langle\boldsymbol{r}_i|\boldsymbol{r}_i\right\rangle.
\end{equation}
This is a positive quantity which can only become zero if $c_i=0$ for $i\neq 0$, and therefore, $\boldsymbol{F}_{(n,\sigma)}$ is proportional to $\boldsymbol{p}^{\eq}$.
The fact that the entropy production can become zero at finite thermodynamic fluxes, means that the Onsager matrix has a zero eigenvalue. This corresponds to the so-called tight coupling limit \cite{van2010many}. 
We shall now briefly show that the adiabatic fluxes go to zero. Using the definition of the fluxes and Eq.~(\ref{Ladft}):
\begin{eqnarray}
{J}^{\ad}_\alpha&=&\frac{1}{\mathcal{T}}\sum_{\beta,n,\sigma,\rho}F_\beta\left\langle \delta\boldsymbol{\gamma}_\alpha|\delta \boldsymbol{\gamma}_\beta\right\rangle (1-\delta_{\sigma,\rho})(-1)^{\delta_{\sigma,s}} \pi na_{(\alpha,n,\sigma)}a_{(\beta,n,\rho)}\nonumber\\&=&\frac{1}{\mathcal{T}}\sum_{n,\sigma,\rho}\left\langle \delta\boldsymbol{\gamma}_\alpha|\boldsymbol{p}^{\eq}\right\rangle (1-\delta_{\sigma,\rho})(-1)^{\delta_{\sigma,s}} \pi na_{(\alpha,n,\sigma)}=0,
\end{eqnarray}
where the last line follows from the definition of the inner product Eq.~(\ref{inpdef}) and the fact that $\sum_n \delta\gamma_{\alpha,n}=0$.
The proof of the vanishing non-adiabatic fluxes, and therefore the total fluxes, is completely analogous. 
Summarising, we conclude that, even if the time-symmetry is explicitly broken, it is not possible to have finite thermodynamic fluxes while maintaining zero entropy production.

\newpage
\bibliographystyle{iopart-num}
\bibliography{main.bib}

\end{document}